# Spectrum of the Supernova Relic Neutrino Background and Evolution of Galaxies

Tomonori Totani[1], Katsuhiko Sato[1,2], and Yuzuru Yoshii[2,3]

[1] Department of Physics, School of Science, the University of Tokyo
7-3-1 Hongo, Bunkyo-ku, Tokyo 113, Japan

[2] Research Center for the Early Universe, School of Science, the University of Tokyo
7-3-1 Hongo, Bunkyo-ku, Tokyo 113, Japan

[3] Institute of Astronomy, Faculty of Science, the University of Tokyo
2-21-1 Osawa, Mitaka, Tokyo 181, Japan

## ABSTRACT

The spectrum of the supernova relic neutrino background (SRN) from collapse-driven supernovae ever occurred in the universe is calculated by using a realistic, time-dependent supernova rate derived from a standard model of galaxy evolution based on the population synthesis method. The SRN spectrum we show here is the most realistic at present, because the largest uncertainty in previous theoretical predictions has come from unrealistic assumptions of the supernova rate so far made. The SRN is one of the targets of the Superkamiokande (SK) detector which will be constructed in a year and the SRN, if at all detected, would provide a new tool to probe the history of supernova explosions in the universe. The expected event rate at the SK is therefore calculated in this paper. Our major results include: (1) the supernova rate is much higher in the early phase of evolution of galaxies and there appears a hump in the SRN spectrum in the low-energy region of $\lesssim 5$ MeV, (2) the SRN flux depends on the Hubble constant ($H_0$) in a way approximately proportional to $H_0^2$ and only weakly on the density parameter of the universe ($\Omega_0$) and a cosmological constant ($\lambda_0$), (3) the uncertainty in the star formation history of spiral galaxies affects the resulting SRN flux by about a factor of 3, and (4) the plausible event rate at the SK is 1.2 yr$^{-1}$ in the observable energy range of 15–40 MeV. Such a low event rate is due mainly to a quite low supernova rate at present which is averaged over the morphological types of galaxies. The most optimistic rate in our model is found to be 4.7 yr$^{-1}$ in the same energy range, and if more events are detected, we will have to reconsider our current understanding of collapse-driven supernovae and evolution of galaxies.

*Subject headings:* cosmology: diffuse radiation—galaxies: evolution—stars: supernovae: general

## 1. Introduction

The water Čerenkov neutrino detector, Superkamiokande (SK) (Nakamura et al. 1994), will be constructed and observations will start in a year. One of the targets of this detector is to detect the supernova relic neutrino background (SRN) which would originate from all collapse-driven supernovae ever occurred in the universe (Bisnovatyi-Kogan & Seidov 1984; Krauss, Glashow, & Schramm 1984; Dar 1985; Woosley, Wilson, & Mayle 1986; Zhang et al. 1988; Hirata 1991; Totsuka 1992; Suzuki 1994; Totani & Sato 1995). Since the emergence of neutrinos from a supernova has been confirmed from both numerical simulations and observations of SN1987A, comparison of the SRN spectrum with observation at the SK will enable us to constrain a supernova rate and therefore provide us valuable information on the star formation history in galaxies. In this regard, in order to foresee a capability of the detector it is worth predicting the SRN spectrum and the resulting event rate at the SK for a given rate of supernovae in various models of the universe, either with or without a cosmological constant. In order to detect the weak SRN signal, various other backgrounds must be removed. Fortunately, there is a quite silent window between 15 and 40 MeV where various backgrounds other than the SRN are strongly suppressed (Hirata 1991, Totsuka 1992). Although it is not certain whether the SRN can be detected even in this window, it is important to predict the most believable flux of the SRN which has only roughly been estimated in a broad range of $1$–$10^2$ cm$^{-2}$s$^{-1}$.

A major source of the uncertainty in previous theoretical predictions has been a supernova rate prescribed in the models. While most of previous papers assume a supernova rate which is constant with time, it is more likely that the past rate is higher than at present as galaxies evolve. The effect of galaxy evolution is considered by Bisnovatyi-Kogan & Seidov (1984), assuming that a supernova rate depends on time or redshift in a manner proportional to $(1+z)^\gamma$, where $z$ is the redshift parameter. This effect is also considered in our previous paper (Totani & Sato 1995, henceforth Paper I), assuming that half of the supernovae have exploded like a burst at a certain redshift and another half have ever occurred at a constant rate. These assumptions are, however, rather artificial and a reasonable model of galaxy evolution should be used for prescribing a realistic, time-dependent supernova rate.

In this paper, using a standard model of evolution of galaxies based on the method of stellar population synthesis (Arimoto & Yoshii 1986, 1987; Arimoto, Yoshii, & Takahara 1992), we calculate a supernova rate as a function of look-back time of galaxies and predict the most realistic SRN spectrum as well as the event rate at the SK for various cosmological models. This model of galaxy evolution has been used to investigate the geometry of the universe from comparison with the number counts of faint galaxies (e.g., Yoshii & Takahara 1988; Fukugita et al. 1990; Yoshii 1993; Yoshii & Peterson 1995) and basically agrees with observations, particularly after the selection effects are taken into account in the theoretical framework (Yoshii 1993). Some variants of the standard model of galaxy evolution are also used to delineate a range of uncertainty in the resulting SRN spectrum arising from different histories of star formation in galaxies.

Throughout this paper, we consider only electron antineutrinos ($\bar{\nu}_e$'s) from collapse-driven supernovae, because the contribution of neutrinos from thermonuclear supernovae to the SRN is negligible and $\bar{\nu}_e$'s are most easily detected in a water Čerenkov detector. In §2, we construct a model of evolution of galaxies, the supernova rate, and neutrino emission from a supernova. Formulations for calculating the SRN spectrum are given in §3. The results are given in §4, and their dependence on models of the universe as well as galaxy evolution is also discussed. After the total (energy-integrated) SRN flux of our results is compared with previous estimates in this section, we make a prediction for the response of the SK detector in §5, and finally summarize the results in §6.

## 2. Model Construction

### 2.1. The Population Synthesis Model of Galaxy Evolution

We summarize here the model of evolution of galaxies used in this paper, following Arimoto & Yoshii (1986, 1987) and Arimoto, Yoshii, & Takahara (1992). It is assumed that a galaxy consists of three components, namely, interstellar gas, stars, and stellar remnants. Time evolution of the mass fraction of gas in a galaxy, $f(t)$, is determined by the differential equation:

$$\frac{df}{dt} = -C(t) + F(t) + A(t) , \tag{1}$$

where $C(t)$ is the star formation rate (SFR), $F(t)$ is the ejection rate of gas from dying stars, $A(t)$ is the gas accretion rate (ACR) from the surrounding region of a galaxy, and $t$ is the time elapsed from the formation of a galaxy, not a cosmic time.

It is generally assumed that the star formation law can be approximated as

$$C(t) = \nu f^n(t), \tag{2}$$

where $\nu$ is the SFR coefficient and has the dimension of (time)$^{-1}$, and $n$ is the power index. A large number of papers constraining the form of SFR from observations show that $n$ is most likely between 1 and 2 for spiral galaxies (e.g., Kennicutt 1989), and we consider two cases of $n = 1$ and $n = 2$ in this paper. For elliptical galaxies, however, the remaining gas would leave the system by a galactic wind driven by a burst of supernova explosions in the early stage of evolution of galaxies. We introduce this effect in the star formation law for elliptical galaxies as follows (Arimoto & Yoshii 1987):

$$C(t) = \begin{cases} \nu f(t) & (t \leq t_{GW}) \\ 0 & (t > t_{GW}) , \end{cases} \tag{3}$$

where $t_{GW}$ is the time when the galactic wind occurs.

The ejection rate, $F(t)$, is expressed as

$$F(t) = \int_{m_l}^{m_u} C(t - t_m) \phi_m R_m dm . \tag{4}$$

Here, $m$ is the stellar mass, $R_m$ is the fraction of ejected mass from a dying star, $\phi_m$ is the initial stellar mass function (IMF), and $m_u$ and $m_l$ are the upper and lower bounds of the stellar mass, respectively. In this paper we adopt $m_l = 0.05 M_\odot$ and $m_u = 60 M_\odot$. The lifetime of a star of mass $m$ is denoted by $t_m$ and is taken as $t_m = 11.7(m/M_\odot)^{-2}$Gyr (Talbot & Arnett 1971). The IMF $\phi_m$ is assumed to be of a power-law form:

$$\phi_m = \frac{(\mu - 1) m_l^{\mu - 1}}{1 - (m_l/m_u)^{\mu - 1}} m^{-\mu} , \tag{5}$$

where $\mu$ is the power index and the function is normalized to unity. The fraction of ejected mass from a dying star is assumed to be

$$R_m = \begin{cases} 0 & (m_l \leq m < m_{rem}) \\ 1 - (m_{rem}/m) & (m_{rem} \leq m < m_u) , \end{cases} \tag{6}$$

with $m_{rem}$ set to be $0.7 M_\odot$ (Talbot & Arnett 1971; Arnett 1978).

Solutions to the above equations depend on the assumptions made for gas accretion from the surrounding region of a galaxy. One is the 'simple', one-zone model, in which there is no accreting gas from the surrounding region, and all the gas contributes to star formation at the initial stage of galaxy evolution, i.e., $A(t) = 0$ and $f(0) = 1$. The other is the 'infall' model, in which there is no gas initially and the gas is supplied gradually from the surrounding region of a galaxy. The form of ACR is assumed as $A(t) = a \exp(-at)$ and $f(0) = 0$ (Arimoto et al. 1992), where $a$ is the ACR coefficient and has the dimension of $(\text{time})^{-1}$. Chemical evolution studies suggest that gas infall plays a decisive role in reproducing the photometric and chemical properties of present-day spiral galaxies (e.g., Matteucci & François 1989; Arimoto & Jablonka 1991). We therefore consider two alternative assumptions such that the accretion is turned on and off for spiral galaxies. On the other hand, for elliptical galaxies, we consider only the simple model without gas infall.

In this paper, we group E and S0 galaxies together and classify spirals into four subclasses like Sab, Sbc, Scd, and Sdm. Evolutionary behaviors of chemical abundances in the gas and stars as well as integrated colors of stars in a galaxy are modeled with the parameters of $\nu$, $t_{GW}$ and $\mu$ for ellipticals, or those of $\nu$, $a$, $n$ and $\mu$ for spirals. These parameters have been fixed through a fit to the observed colors of galaxies. For spirals from Sab to Sdm, the adopted values of $\nu$ and $a$ for $n = 1$ or 2, together with Salpeter's IMF ($\mu = 1.35$), are the same as those in Arimoto et al. (1992). For ellipticals, the adopted values of $\nu$ and a corresponding $t_{GW}$ are those for the lower, standard, and higher SFR models in Arimoto & Yoshii (1987); the lower and higher bounds of $\nu$ correspond to an intrinsic dispersion of SFR, which can account for the observed scatter along the mean relationship on the color-magnitude (CM) diagram. We note that a somewhat flat IMF with $\mu = 0.95$ well reproduces the observed optical through infrared colors of ellipticals (Arimoto & Yoshii 1987) and we adopt this value of $\mu$ for E/S0-type galaxies.

In this paper, we consider four cases (simple/infall and $n = 1$ or 2) for each of Sab, Sbc, Scd and Sdm, and three cases (lower, standard, and higher SFRs) for E/S0-type galaxies; these models are considered to cover the uncertainty of the evolution model of galaxies. Except when the uncertainty of galaxy evolution is concerned, we use the S1 model to represent the evolution of spiral galaxies, and the standard SFR model for E/S0-type galaxies; here and hereafter, for simplicity, we refer S1 or S2 to the simple model with $n = 1$ or 2, and similarly I1 or I2 to the infall model with $n = 1$ or 2.

## 2.2. Comparison with the Number Counts of Faint Galaxies

In a number of papers trying to determine the geometry of the universe from the number counts of faint galaxies, the above model of galaxy evolution has been used and basically shown to agree with observations in some different wave bands (e.g., Yoshii & Takahara 1988; Fukugita et al. 1990; Yoshii 1993; Yoshii & Peterson 1995), while the values of the density parameter of the universe ($\Omega_0$) and a cosmological constant ($\lambda_0$) have not yet been settled. In this section, we compare the predicted number counts of faint galaxies with the observations to show the validity of using the model constructed in the previous section.

The $K$-band data of number counts are those compiled in Yoshii & Peterson (1995) with the addition of some new observations (Djorgovski et al. 1995; McLeod et al. 1995; Aragon-Salamanca & Ellis 1994). Among all these surveys, Djorgovski et al.'s survey uses the faintest surface brightness threshold for image detection, thus providing the faintest number count data. Given a fixed threshold, the selection effects become quite significant near the survey limit (Yoshii 1993) and these effects should be taken into account in the predictions to be compared with the uncorrected counts in the faint end. The observational conditions used to predict the uncorrected counts are matched with those actually used by Djorgovski et al. (1995), that is, the 1.5″ aperture magnitude with $S_L = 24$ mag/arcsec$^2$ and the typical seeing of 1″ FWHM, where $S_L$ is the threshold of surface brightness. For the minimum angular diameter, we adopt $D_{min} = 1''$. The manner of including the selection effects is described in Yoshii (1993).

Figure 1 shows the observations of the $K$-band number counts and the predicted curves from the standard model of galaxy evolution (the S1 model is used for evolution of spiral galaxies, and the standard SFR model for ellipticals). The data are taken from various surveys with different $S_L$'s and the error bars represent the Poissonian error. The data with brighter $S_L$'s should be corrected for undetected images when plotted in the same figure with Djorgovski et al.'s data with the faintest $S_L$. Therefore, except for Djorgovski et al.'s uncorrected counts to which the predictions are compared, shown are the corrected counts from all the other surveys with brighter $S_L$'s.

Three different cosmological models are considered: a low density, flat universe with $\Omega_0 = 0.2, \lambda_0 = 0.8$, and $h = 0.8$ (thick line), a low density, open universe with $\Omega_0 = 0.2, \lambda_0 = 0.0$, and $h = 0.8$ (dashed line), and an Einstein-de Sitter universe with $\Omega_0 = 1.0, \lambda_0 = 0.0$, and $h = 0.5$ (dot-dashed line), where $h = H_0/(100\mathrm{km/s/Mpc})$. Note that changing the Hubble constant hardly affects the theoretical predictions of number counts. The parameters of the luminosity function are those in Efstathiou, Ellis & Peterson (1988). This figure shows that the model predictions are well consistent with observations, while the model of the universe cannot be determined strictly. We calculate the SRN spectrum with these three cosmological models.

Figure 2 shows the observations of the number counts and theoretical predictions of the model of galaxy evolution for only a low density, flat universe ($\Omega_0 = 0.2, \lambda_0 = 0.8, h = 0.8$) with the four different models of evolution of spirals; S1 (thick line), S2 (short dashed line), I1 (dot and short-dashed line), and I2 (dot and long-dashed line). The effects of luminosity evolution are least in redder passbands because the steady accumulation of long-lived, low-mass stars in galaxies makes the evolution of the red light secular and weak (Bruzual 1983; Arimoto & Yoshii 1986), and all of these four models are well consistent with the observations of the number counts of faint galaxies.

### 2.3. Supernova Rate

We have constructed the model of evolution of galaxies from which the supernova rate is obtained. The neutrinos of our concern are emitted from collapse-driven supernovae having massive progenitor stars, and their rate per unit mass of a galaxy is expressed as

$$R_{SN}(t) = \begin{cases} 0 & (0 \leq t < t_{m_u}) \\ \int_{\max(m_t, m_{SNl})}^{m_u} \phi_m m^{-1} C(t - t_m) dm & (t \geq t_{m_u}) \end{cases}, \qquad (7)$$

where $m_t$ is the mass for which $t_m = t$, and $m_{SNl}$ is the critical mass above which all stars end their life with collapse-driven supernova explosions; we adopt $m_{SNl} = 8 M_\odot$ throughout this paper. The SFR per unit mass of a galaxy, $C(t)$, is numerically calculated from equations in §2.1.

The supernova rates thus calculated are shown in Figs. 3 to 5, in units of yr$^{-1}$ in a galaxy with the typical mass of $10^{11} M_\odot$. Figure 3 shows by the dashed lines the supernova rates for five types of galaxies using the S1 model for spirals and the standard SFR model for ellipticals. The supernova rate averaged over the types with a weight of $a_k(M/L_B)_k$ is shown by the thick line, where $a_k$ is the relative proportion of the $k$-th type galaxies (Pence 1976) and $(M/L_B)_k$ is their $B$-band mass-to-luminosity ratio (Faber & Gallagher 1979). The values of $a_k$ and $(M/L_B)_k$ are tabulated in Table 1. We note that the 'mass' of a galaxy in this paper represents the mass of the ordinary matter, i.e., field stars, remnants, and interstellar gas which is used in star formation. The mass-to-luminosity ratio is that within the Holmberg radius, within which the luminous and ordinary matter is considered to be dominant. Figure 4 shows the weighted average rates with the four different models of S1, S2, I1, and I2 for spirals provided that the standard SFR model for ellipticals is used in common. Figure 5 shows the weighted average rates with the three different models (lower, standard, and higher SFR) for ellipticals provided that the S1 model for spirals is used in common.

In all the cases, supernovae most dominantly occur in the early phase of galaxy evolution ($\lesssim 1$ Gyr) in E/S0-type galaxies, and then in the later phase ($\gtrsim 1$ Gyr) in spirals. This is a general trend largely irrespective of different assumptions made for galaxy evolution, as clearly seen in Figs. 3 to 5. It should be noted that the weighted average rate in the early phase is about two orders of magnitude higher when compared to that in the recent past, and more than the half of a total number of supernovae have exploded during the initial 1 Gyr; this feature affects strongly the shape of the resulting SRN spectrum.

Now we compare our predicted average rate of collapse-driven supernovae with observations. There are various observational estimates of the supernova rate, and they are scattered with a mean around one collapse-driven supernova in 50 years in our Galaxy (Tammann et al. 1994). In general, our Galaxy is considered to be an Sbc galaxy with an age of 10–15 Gyrs, having a total mass of about $10^{11} M_\odot$. With these typical values of age and mass, our predicted supernova rate for Sbc spans from one in 50 years to one in 100 years (Fig.3) in fairly good agreement with the observational estimate.

We also examine whether or not our model meets the nucleosynthesis requirement. Since the oxygen is synthesized in the interior of massive stars and intermediate mass nuclei burn away to create iron in thermonuclear supernovae, the amount of oxygen is related directly to the time-integrated number of collapse-driven supernovae. Figure 6 shows the time-integrated number of collapse-driven supernovae in a galaxy with the mass of $10^{11} M_\odot$ as a function of time from the formation of the galaxy; results for individual types of galaxies are shown by the dashed lines and those for the weighted average by the thick line. Here, as in Figure 3, the S1 model is used for spirals and the standard SFR model for ellipticals, but the results are hardly changed by other choices of evolution models. Taking the mass fraction of oxygen in the Galactic disk as about $8.5 \times 10^{-3}$ from observations of galactic HII regions (e.g., Díaz 1989), the total mass of oxygen in a galaxy of $10^{11} M_\odot$ is estimated as $8.5 \times 10^8 M_\odot$. Arnett, Schramm, & Truran (1989) showed that about $10^9$ collapse-driven supernovae should be responsible for the oxygen enrichment to that amount in our Galaxy. Our predicted number for an Sbc galaxy is about $10^9$ at present, which agrees with that required from the nucleosynthesis arguments. This agreement makes the evolution models used in this paper more believable, because the models have been contrived to reproduce the observed colors of galaxies, but not the cumulative number of supernovae in galaxies.

### 2.4. Electron Antineutrino Emission from a Supernova

Emission of electron antineutrinos ($\bar{\nu}_e$'s) from a supernova is treated in the same way as Paper I. We assume that the energy distribution of the neutrinos obeys the Fermi-Dirac distribution without chemical potential. Numerical calculations of Woosley et al. (1986) for gravitational collapses of massive stars with 10, 15, and $25 M_\odot$ are used to represent the total energy and temperature of $\bar{\nu}_e$'s from a supernova in the three intervals of stellar mass such as 8–$12.5 M_\odot$, $12.5$–$20 M_\odot$, and 20–$60 M_\odot$, respectively. Contrary to Paper I where the number of stars in these three intervals is assumed to be constant with time, the number of stars changes with time in a way consistent with the model of galaxy evolution considered in this paper. Following the above approximation, we denote a differential number of emitted neutrinos from a supernova as $dN_\nu(q)/dq$ where $q$ is the neutrino energy.

## 3. Formulations

The differential number flux of SRN is expressed as $dF_\nu(q)/dq = c \, dn_\nu(q)/dq$ where $dn_\nu(q)/dq$ is the present number density of SRN per unit energy of neutrinos, and $c$ is the speed of light. The neutrinos emitted in the interval of cosmic time between $t$ and $t + dt$ contribute to $dn_\nu(q)$ as follows:

$$dn_\nu(q) = \rho_G(t) R_{SN}(t - t_F) dt \frac{dN_\nu\{(1+z)q\}}{dq} \{(1+z)dq\}(1+z)^{-3} , \qquad (8)$$

where $\rho_G(t)$ is the mass density of galaxies in the universe, $t_F$ is the cosmic time when galaxies were formed, $z$ is the redshift parameter, and $R_{SN}(t - t_F)$ is the supernova rate at the cosmic time $t$ per unit mass of galaxies. Note that the definition of $R_{SN}$ in this paper is different from that in Paper I. The differential number of emitted neutrinos from a supernovae, $dN_\nu(q)/dq$, is given in §2.4. The factor $(1+z)^{-3}$ comes from the expansion of the universe, and using $\rho_G(t) = \rho_G(t_0)(1+z)^3$, equation (8) reduces to

$$\frac{dn_\nu(q)}{dq} = \rho_G(t_0) R_{SN}(t - t_F) \frac{dN_\nu\{(1+z)q\}}{dq}(1+z) dt . \qquad (9)$$

In order to proceed further, the quantity $\rho_G(t_0) R_{SN}(t - t_F)$ is given by

$$\rho_G(t_0) R_{SN}(t - t_F) = \sum_k \int dL L \psi_k(L) \left(\frac{M}{L}\right)_k R_k(t - t_F) , \qquad (10)$$

where $\psi_k(L)$, $(M/L)_k$, and $R_k(t - t_F)$ are the luminosity function, the mass-to-luminosity ratio, and the supernova rate for the $k$-th type galaxies, respectively, and the subscript $k$ runs over five types of galaxies (E/S0, Sab, Sbc, Scd, Sdm). By definition, $\psi_k(L)$ is the number density of the $k$-th type galaxies having the absolute luminosity $L$. We assume that the shape of the luminosity function does not depend on the type, yielding $\psi_k(L) = a_k \psi(L)$ where $a_k$ is the relative proportion of each type and $\psi(L)$ is the luminosity function for the composite of all types of galaxies. Furthermore, we assume that the mass-to-luminosity ratio does not depend on the luminosity of galaxies. Accordingly, equation (9) becomes

$$\frac{dn_\nu(q)}{dq} = \mathcal{L} \left[\sum_k a_k \left(\frac{M}{L}\right)_k R_k(t - t_F)\right] \frac{dN_\nu\{(1+z)q\}}{dq}(1+z) dt , \qquad (11)$$

where $\mathcal{L} = \int dL L \psi(L)$ is the luminosity density of galaxies. Integrating over $z$ and using a relation $dF_\nu(q)/dq = c \, dn_\nu(q)/dq$, we finally obtain the differential number flux of SRN:

$$\frac{dF_\nu(q)}{dq} = c\mathcal{L} \int_0^{z_F} \left[\sum_k a_k \left(\frac{M}{L}\right)_k R_k(t - t_F)\right] \frac{dN_\nu\{(1+z)q\}}{dq}(1+z) \left|\frac{dt}{dz}\right| dz , \qquad (12)$$

with

$$\frac{dz}{dt} = -H_0(1+z)\sqrt{(1+\Omega_0 z)(1+z)^2 - \lambda_0(2z + z^2)} , \qquad (13)$$

where $z_F$ is the formation redshift of galaxies, and $\lambda_0$ is the normalized cosmological constant; $\lambda_0 = (c^2 \Lambda)/(3H_0^2)$. The time $t - t_F$ is related to $z$ via $t - t_F = \int_z^{z_F} |dt/dz| dz$. Throughout this paper, following Efstathiou, Ellis & Peterson (1988), we adopt the $B$-band luminosity density of $\mathcal{L}_B = 1.93^{+0.8}_{-0.6} \times 10^8 h \, L_{B\odot} \, \text{Mpc}^{-3}$, where $h = H_0/(100 \text{km}\,\text{s}^{-1} \text{Mpc}^{-1})$. The recent results of the luminosity function (Loveday et al. 1992; Marzke et al. 1994) are also consistent with this value of $\mathcal{L}_B$. The adopted values of $a_k$ (Pence 1976) and $(M/L_B)_k$ (Faber & Gallagher 1979) are summarized in Table 1. With the above values of $a_k, (M/L)_k$, and $\mathcal{L}$, the contribution of the mass of galaxies to the density parameter of the universe, $\Omega_G$, becomes about 0.7%. We set $z_F = 5$, except for otherwise studying the effect of changing $z_F$ in §4.2.

In most of previous papers including Paper I, the absolute number of supernovae in the model is determined by a fit to the observation of the present supernova rate in our Galaxy, which is quite uncertain. In this paper, instead of using the present supernova rate, however, we use the observations of the luminosity density of galaxies in the universe and the mass-to-luminosity ratio of galaxies, and then derive the absolute number of supernovae in the model.

## 4. Supernova Relic Neutrino Background (SRN)

### 4.1. SRN Spectrum in Some Cosmological Models

We calculate the SRN spectrum in some models of the universe, using the S1 model for evolution of spirals and the standard SFR model for ellipticals. We use a low-density, flat universe with $\Omega_0 = 0.2$, $\lambda_0 = 0.8$, and $h = 0.8$ as a standard cosmological model. The SRN spectrum for this standard cosmological model and two variants with $h = 0.5$ and 1.0 are shown in Figure 7, by the thick lines. The spectra for a low-density, open universe with $\Omega_0 = 0.2$, $\lambda_0 = 0.0$, and $h = 0.8$ (dashed line) and an Einstein-de Sitter universe with $\Omega_0 = 1.0$, $\lambda_0 = 0.0$, and $h = 0.5$ (dot-dashed line) are also shown in this figure.

For a purpose of comparison, the SRN spectrum with a constant supernova rate (from Paper I) is also shown by the dotted line for a standard cosmological model ($\Omega_0 = 0.2$, $\lambda_0 = 0.8$, $h = 0.8$). In order not to violate the nucleosynthesis requirement, the value of the constant supernova rate ($1.6 \times 10^{-3} \text{yr}^{-1} \text{Mpc}^{-3}$) is chosen such that the total SRN flux $\int (dF_\nu/dq) dq$, or equivalently the total number of supernovae ever occurred, is the same as that for the time-dependent supernova rate considered in this paper. Thus, from these two SRN spectra we clearly see how the effect of galactic evolution modifies the resulting SRN spectrum.

In all SRN spectra, there appears a hump in the low-energy region of $\lesssim 5$ MeV. This feature is similar even to the constant+burst model in Paper I. The reason is because a very large number of neutrinos emitted from supernova bursts in the early phase of galaxy evolution are strongly redshifted to populate the low-energy part of the spectrum. On the other hand, in the high-energy region of $\gtrsim 10$ MeV, most of neutrinos are from supernovae which occurred at a smaller rate in the recent past.

The SRN flux depends most significantly on the Hubble constant among the cosmological parameters, and it approximately scales as $dF_\nu/dq \propto H_0^2$. This is understood as follows: The SRN flux is proportional to the mass density of galaxies in the universe, $\mathcal{M}$, and its observational estimates are generally proportional to $H_0^2$ like $\mathcal{M} = \mathcal{L} \times (M/L) \propto H_0 \times H_0$. Furthermore, the total number of supernovae and therefore the total SRN flux apparently increase with increasing the age of galaxies, or equivalently with decreasing $H_0$ because the age is proportional to $H_0^{-1}$. This effect is, however, very small; most of the supernovae have exploded within the initial 1 Gyr from the formation of galaxies, so that the total number of supernovae does not significantly change with the age of galaxies (Fig.6). On the other hand, shortening of the age leads to a higher present supernova rate (Fig.3), therefore the SRN flux is enhanced in the high-energy region ($\gtrsim 10$ MeV) where neutrinos are easily detected in the SK detector. As a result, the event rate at the SK scales as $H_0^2$ or a power of $H_0$ with a little higher index.

Within a considered range of $\Omega_0 = 0.2$–1, $\lambda_0 = 0$–1, and $h = 0.5$–1, the SRN spectrum is less sensitive to $\Omega_0$ or $\lambda_0$ when compared to $H_0$. Smaller $\Omega_0$ or larger $\lambda_0$ decreases the SRN flux in the high-energy region, since these effects equally prolong the age of galaxies and therefore decrease the supernova rate in the recent past.

### 4.2. Uncertainty from Models of Galaxy Evolution

We estimate a range of uncertainty of the SRN spectrum arising from the uncertainty associated with the model of galaxy evolution. Figure 8 shows the SRN spectra for the S1, S2, I1, and I2 models for spirals provided that the standard SFR model for ellipticals and a standard cosmological model ($\Omega_0 = 0.2, \lambda_0 = 0.8, h = 0.8$) are used in common. In order from S2 through S1 and then I2 to I1, supernovae occur more recently (Fig.4) and the SRN flux becomes higher in the high-energy region of $\gtrsim 10$ MeV, and the S1 model gives an intermediate flux. The SRN flux for the I1 model is about a factor of 2 or 3 higher than that for the S2 model, and this could be regarded as a maximum uncertainty from evolution models of spiral galaxies. On the other hand, the difference among the lower, standard, and higher SFRs for ellipticals hardly changes the resulting SRN spectrum.

Finally, we examine how the SRN spectrum depends on the epoch of galaxy formation, $z_F$. Figure 9 shows the spectra for $z_F =3$, 4, and 5, using the S1 and standard SFR model of galaxy evolution and a standard cosmological model ($\Omega_0 = 0.2, \lambda_0 = 0.8, h = 0.8$). Apparently, the dependence on $z_F$ is small compared to that from the models of spiral galaxies. It is almost negligible above 15 MeV and changes the flux at most by a factor of 2 at about 10 MeV.

### 4.3. Total SRN Flux

We calculate the energy-integrated SRN flux in both the whole energy range and observable energy range, and compare them with previous estimates. Woosley et al. (1986) estimated the total SRN flux in the whole energy range using some different methods, and their estimates spread over a range of 1–250 cm$^{-2}$s$^{-1}$. The total SRN flux in our 'standard' model ($\Omega_0 = 0.2, \lambda_0 = 0.8, h = 0.8, z_F = 5$, the S1 model for spirals, the standard SFR model for ellipticals, and $\mathcal{L}_B = 1.93 \times 10^8 h\ L_{B\odot}\mathrm{Mpc}^{-3}$) gives an moderate value of 44 cm$^{-2}$s$^{-1}$. However, since the SRN cannot be observed in the whole energy range, we have to limit the range to avoid various other backgrounds. In the energy range of 15–40 MeV, the other backgrounds are quiet (Hirata 1991; Totsuka 1992, and see the next section for more detailed discussion), and we can observe the SRN neutrinos through this window, however, only the neutrinos from supernovae which occurred in the recent past can contribute to the flux in this range. The SRN flux of our standard model in this range is as small as 0.83 cm$^{-2}$s$^{-1}$, because the supernova rate which is averaged over morphological types of galaxies is only less than 0.01 yr$^{-1}$ in a galaxy with the mass of $10^{11}M_\odot$ in the recent past, about 10 Gyr after the formation of galaxies (Fig 3). The adopted luminosity density (Efstathiou et al. 1988) corresponds to the number density of galaxies, $n_G = 3.9 \times 10^{-2} h^3$ Mpc$^{-3}$, when all the galaxies are assumed to have a mass of $10^{11} M_\odot$, then the supernova rate per unit volume at the present time should be less than $\sim 3.9 \times 10^{-4} h^3$yr$^{-1}$Mpc$^{-3}$. This value is close to the conservative estimate by Woosley et al. (1986) of $2.2 \times 10^{-4} h^3$yr$^{-1}$Mpc$^{-3}$, which gives the total SRN flux of only about 1 cm$^{-2}$s$^{-1}$ if the rate is assumed to be constant throughout the history of galaxy evolution. Most of the total flux of 44 cm$^{-2}$s$^{-1}$ consists of the neutrinos from supernovae in the early phase of galaxy evolution, which cannot be observed. In estimating the SRN flux in Paper I, we used more optimistic values of the supernova rate of 0.1 yr$^{-1}$ in a galaxy and the number density of galaxies $n_G = 1.6 \times 10^{-1} h^3$ Mpc$^{-3}$ which do not agree with current estimates and therefore unlikely. Consequently, the event rate at the SK should also be very small, which is the subject of the next section.

## 5. Response of the Superkamiokande Detector

In this section, we calculate the expected event rate at the water Čerenkov neutrino detector, Superkamiokande (SK) (Nakamura et al. 1994), and discuss the rate obtained. The procedure of converting the SRN flux to an event rate at the SK is described in Paper I. We consider only the reaction $\bar{\nu}_e p \to e^+ n$ because the cross section ($9.72 \times 10^{-44} E_e p_e$cm$^2$, where $E_e$ and $p_e$ are the energy and momentum of the recoil positrons measured in MeV, respectively) is much larger than those of the other reactions. The detection efficiency of the SK, which has been provided by Totsuka (1994), is taken into account in our calculations (see also Paper I).

We calculate the expected event rate as a function of positron kinetic energy using the S1 and standard SFR model of galaxy evolution for various cosmological parameters (Fig.10), and also varying the evolution model of spiral galaxies for a standard cosmological model ($\Omega_0 = 0.2$, $\lambda_0 = 0.8$, and $h = 0.8$) (Fig.11). These figures correspond to Figures 7 and 8, respectively, where the SRN spectra are shown. The common feature in all the models is that the event rate is peaked at about 5 MeV with a broad plateau over a range of 5–10 MeV, and then decreases rapidly towards higher energies. The event rate does not depend strongly on the assumptions of the SFR of elliptical galaxies or the epoch of galaxy formation, $z_F$. The event rate of a model with a constant supernova rate is also shown in Figure 10, for which the constant rate is normalized such that the total SRN flux is the same as that of the model with $\Omega_0 = 0.2$, $\lambda_0 = 0.8$, and $h = 0.8$ (thick, middle line).

Now we consider various backgrounds which may be an obstacle to the SRN detection. They are solar and atmospheric neutrinos, electron antineutrinos from nuclear reactors, and non-neutrino terrestrial backgrounds ($\gamma$ rays and $\beta$ decays from unstable nuclei produced by cosmic ray muons or spallation products). Below 10 MeV, the SRN will be completely hidden by the reactor $\bar{\nu}_e$'s (Lagage 1985, 1995). In the energy range of 10–15 MeV or 10–20 MeV, the main sources of background events are solar neutrinos and non-neutrino terrestrial backgrounds, however the background event rate associated with these sources drops considerably above 20 MeV(Hirata 1991; Totsuka 1992). Atmospheric neutrinos will dominate the SRN above the energy of about 30–40 MeV (Gaisser et al. 1988; Barr et al. 1989). Therefore, we calculate the energy-integrated event rate of the SRN in the range of kinetic energy of positrons for 15–40 MeV, and also in the more conservative energy range of 20–30 MeV. The energy-integrated event rates in these energy ranges for various models of galaxy evolution and the universe, which are considered in this paper, are tabulated in Table 2.

From the reasons discussed in §4.3, the event rate in the observable range of 15–40 MeV must be quite small. When we use the 'standard' model ($\Omega_0 = 0.2$, $\lambda_0 = 0.8$, $h = 0.8$, $z_F = 5$, the S1 model for spirals, the standard SFR model for ellipticals, and $\mathcal{L}_B = 1.93 \times 10^8 h\ L_{B\odot} \text{Mpc}^{-3}$), the expected event rate at the SK is 1.2 yr$^{-1}$ in such an energy range. The much higher rate of supernova explosions in the early phase of elliptical galaxies causes almost no enhancement in the event rate above 15 MeV, since the neutrinos emitted from such supernovae are strongly redshifted and fall below the observable energy range.

The event rate, like the SRN spectrum, shows a rather strong dependence on the evolution model of spiral galaxies, the Hubble constant, and the luminosity density of galaxies. It shows only a weak dependence on $\Omega_0, \lambda_0$, and the epoch of galaxy formation, $z_F$. Considering all the factors that enhance the present supernova rate, we are able to estimate the 'optimistic' event rate. For example, using $H_0 = 100 \text{km s}^{-1} \text{Mpc}^{-1}$, $z_F = 3$, the I1 model for spirals, and the upper end of the $B$ band luminosity density $\mathcal{L}_B = (1.93 + 0.8) \times 10^8 h\ L_{B\odot} \text{Mpc}^{-3}$ (Efstathiou et al. 1988), the event rate at the SK becomes 4.7 yr$^{-1}$ after integrating over the observable energy range of 15–40 MeV. This value is regarded as an upper bound of the expected event rate, and if more events are observed, we will have to reconsider a current model of evolution of galaxies from which the time-dependent supernova rate is prescribed.

## 6. Summary and Conclusions

We have calculated the SRN spectrum and also the expected event rate at the SK detector by using a realistic, time-dependent supernova rate from a reasonable model of galaxy evolution based on the population synthesis method. This model of galaxy evolution has been used to investigate the geometry of the universe from comparison with the number counts of faint galaxies, and the predicted number counts are well consistent with the recent observations. In our analysis the absolute SRN flux in the model is determined from the luminosity function and the mass-to-luminosity ratio of nearby galaxies, but not from the present supernova rate of our Galaxy which is subject to a large margin of uncertainty.

In the model we used, the supernova rate is much higher in the early phase of elliptical galaxies, and more than half of total supernovae explode during the initial 1 Gyr from the formation of galaxies. Thereafter, the supernova rate in spiral galaxies becomes dominant and the total number of supernovae until present is consistent with the nucleosynthesis requirement. A constant rate of supernova explosions, which was assumed in most of previous papers, is not justified from a viewpoint of galaxy evolution, and therefore the SRN spectrum we show here is the most realistic at the present time.

Supernovae produced at a very high rate in the early phase of ellipticals are responsible for the appearance of a hump in the low-energy part ($\lesssim$ 5 MeV) of the SRN spectrum, because the energy of neutrinos from those supernovae is degraded by the cosmological redshift effect. This energy degradation makes early neutrinos unobservable and the high supernova rate causes almost no enhancement in the event rate in the observable energy range of 15–40 MeV at the SK.

We used a low-density, flat universe with a non-zero $\Lambda$ as a standard cosmological model ($\Omega_0=0.2$, $\lambda_0 = 0.2$, and $h = 0.8$), and investigated the SRN spectrum for other cosmological models relative to the standard model. The SRN flux strongly depends on $H_0$ as roughly proportional to $H_0^2$. The SRN flux changes with $\Omega_0$ and $\Lambda$ in a way that smaller $\Omega_0$ and/or larger $\Lambda$ suppresses the flux in the high-energy region ($\gtrsim$ 10MeV), however, these changes are rather small. The shape of the SRN spectrum is almost insensitive to either of the above three parameters.

We investigated the uncertainty of the SRN spectrum from the model of galaxy evolution. The simple, one-zone and $n = 1$ (S1) model, which we use as a standard evolution model of spiral galaxies, gives an intermediate SRN flux among the four different models of spiral galaxies considered in this paper (for explanation of the models see §2.1). The difference between the highest flux from the I1 model and the lowest flux from the S2 model is within a factor of 3. In the case of elliptical galaxies, however, the different SFRs hardly change the SRN spectrum in either shape or intensity. The effect of changing $z_F$ on the SRN is almost negligible above 15 MeV, and from $z_F = 3$ to 5 the event rate at the SK changes only by a factor of 1.3.

The total SRN flux in the whole energy range becomes 44 cm$^{-2}$s$^{-1}$ in our 'standard' model ($\Omega_0 = 0.2$, $\lambda_0 = 0.8$, $h = 0.8$, $z_F = 5$, the S1 model for spirals, the standard SFR model for ellipticals, and $\mathcal{L}_B = 1.93 \times 10^8 h \, L_{B\odot} \mathrm{Mpc}^{-3}$), which is a moderate value compared with the previous estimates, however, the SRN flux in the observable energy range of 15–40 MeV is as small as 0.83 cm$^{-2}$s$^{-1}$ corresponding to a low end of the SRN flux estimated before. This comes from the very small present supernova rate which is averaged over morphological types of galaxies with the weight of the mass-to-luminosity ratio. This small flux directly leads to a low event rate, and makes the SRN detection at the SK more difficult.

The event rates at the SK detector are calculated for the various models considered in this paper (Table 2), and in the case of the 'standard' model, the expected rate is as small as 1.2 yr$^{-1}$ in the observable range of 15–40 MeV. The event rate depends mainly on the evolution model of spiral galaxies, the Hubble constant, and the luminosity density of galaxies, while only weakly on the geometry of the universe (i.e., on $\Omega_0$ and $\lambda_0$) and the epoch of galaxy formation. These properties are in contrast with those of the number counts of faint galaxies, and the SRN observation may be complementary to the number counting of faint galaxies, if the SRN is detected. From the dependence of the flux on the Hubble constant, the SRN signal would provide a new test for measuring the Hubble constant. However, because of presumably low event rates, such a test may not be possible in practice.

Changing the model parameters within the allowable range in a way of increasing the event rate ($h \to 1$, $z_F \to 3$, S1 $\to$ I1, and $\mathcal{L}_B \to (1.93 + 0.8) \times 10^8 h \, L_{B\odot} \mathrm{Mpc}^{-3}$), we predict the 'most optimistic' event rate at the SK, which is 4.7 yr$^{-1}$ in the energy range of 15–40 MeV. The very small rate of expected SRN events at the SK implies the difficulty of gaining some insight on the star formation history in galaxies from the SRN observation in the SK experiment. More events exceeding 4.7 yr$^{-1}$, if detected, may enforce a serious revision in our current understanding of collapse-driven supernovae and evolution of galaxies.

We would like to thank Y. Totsuka for valuable discussion and useful information about the Superkamiokande detector. We would also like to thank P.O. Lagage for the updated data of the reactor $\bar{\nu}_e$ flux. We also appreciate a critical reading of this manuscript by L.M. Krauss. This work has been supported in part by the Grant-in-Aid for COE Reaearch (07CE2002) of the Ministry of Education, Science, and Culture in Japan.

TABLE 1

Relative Proportions and Mass-to-Luminosity Ratios of Galaxy Types

| Quantity | E/S0 | Sab | Sbc | Scd | Sdm | References |
|---|---|---|---|---|---|---|
| Relative Proportion... | 0.215 | 0.185 | 0.160 | 0.275 | 0.165 | 1 |
| $(M/L_B)$ [a] | 16.2 | 13.0 | 9.4 | 7.8 | 3.4 | 2 |

[a] Measured in the $B$ band within the Holmberg radius; in units of $h \times (M/L_B)_\odot$ where $h \equiv H_0/100 \mathrm{km\,s^{-1}Mpc^{-1}}$.

References.— (1) Pence 1976; (2) Faber & Gallagher 1979.

TABLE 2

Expected Event Rates at the Superkamiokande Detector

| | | | | | | Event Rate (yr$^{-1}$) | |
|---|---|---|---|---|---|---|---|
| $\Omega_0$ | $\lambda_0$ | h | $z_F$ | S-type evolution [a] | $\mathcal{L}_B$ [b] | 15–40 MeV [c] | 20–30 MeV [c] |
| 1.0 | 0.0 | 0.5 | 5 | S1 | 1.93 | 0.63 | 0.28 |
| 0.2 | 0.0 | 0.8 | 5 | S1 | 1.93 | 1.4 | 0.62 |
| 0.2 | 0.8 | 0.5/0.8/1.0 | 5 | S1 | 1.93 | 0.40/1.2/1.9 | 0.17/0.51/0.83 |
| 0.2 | 0.8 | 0.8 | 3/4/5 | S1 | 1.93 | 1.4/1.3/1.2 | 0.66/0.54/0.51 |
| 0.2 | 0.8 | 0.8 | 5 | S1/S2/I1/I2 | 1.93 | 1.2/0.78/1.8/1.5 | 0.51/0.34/0.79/0.65 |
| 0.2 | 0.8 | 1.0 | 3 | I1 | 1.93+0.8 | 4.7 | 2.1 |

[a] The standard SFR model is used in common to represent the E/S0-type evolution.

[b] The luminosity density of galaxies in units of $10^8 h \times L_{B\odot}$ Mpc$^{-3}$ where $h \equiv H_0/100\mathrm{km\,s^{-1}Mpc^{-1}}$.

[c] Kinetic energy of recoil positrons.

Note.— We refer the 'standard' model to the case with ($\Omega_0$, $\lambda_0$, h, $z_F$, S-type evolution, $\mathcal{L}_B$)= (0.2, 0.8, 0.8, 5, S1, 1.93), and the 'most optimistic' model to the case given in the last row with (0.2, 0.8, 1.0, 3, I1, 1.93+0.8).

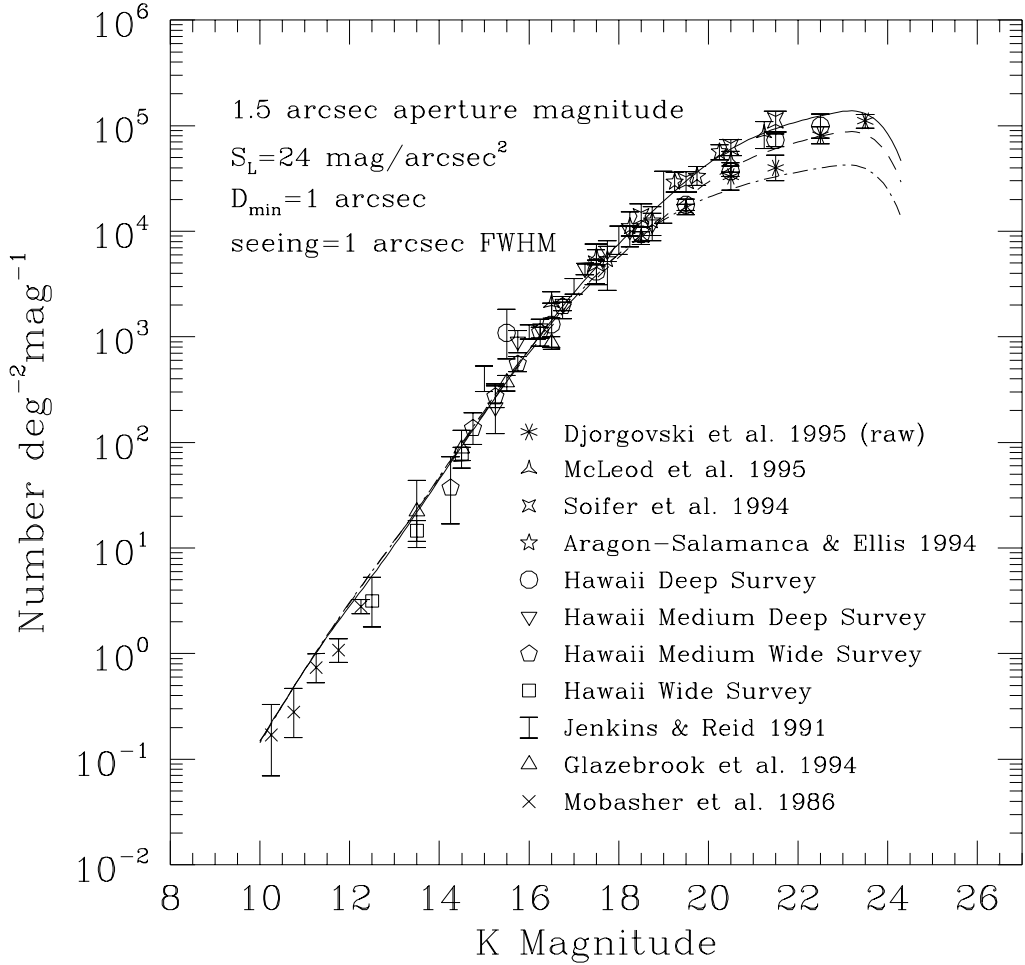

Fig. 1.— The differential number counts of galaxies in the infrared K band in a 1 mag interval in a unit area of 1 deg$^2$, calculated from the model of galaxy evolution for three different cosmological models. The S1 model is used to represent evolution of spiral galaxies, and the standard SFR model for elliptical galaxies (see text). The symbols represent the observed counts taken from various references. The data from the Hawaii observations are taken from Gardner, Cowie & Wainscoat (1993), and references of the other observations are given in the lower right corner of the figure. The error bars stand for the Poissonian error. The selection effects are included in the model curves, using the observational conditions of Djorgovski et al.'s survey which are given on the upper left corner of the figure. The thick line corresponds to a low density, flat universe with $\Omega_0 = 0.2$, $\lambda_0 = 0.8$, and $h = 0.8$, the dashed line to a low density, open universe with $\Omega_0 = 0.2$, $\lambda_0 = 0.0$, and $h = 0.8$, and the dot-dashed line to an Einstein-de Sitter universe with $\Omega_0 = 1.0$, $\lambda_0 = 0.0$, and $h = 0.5$.

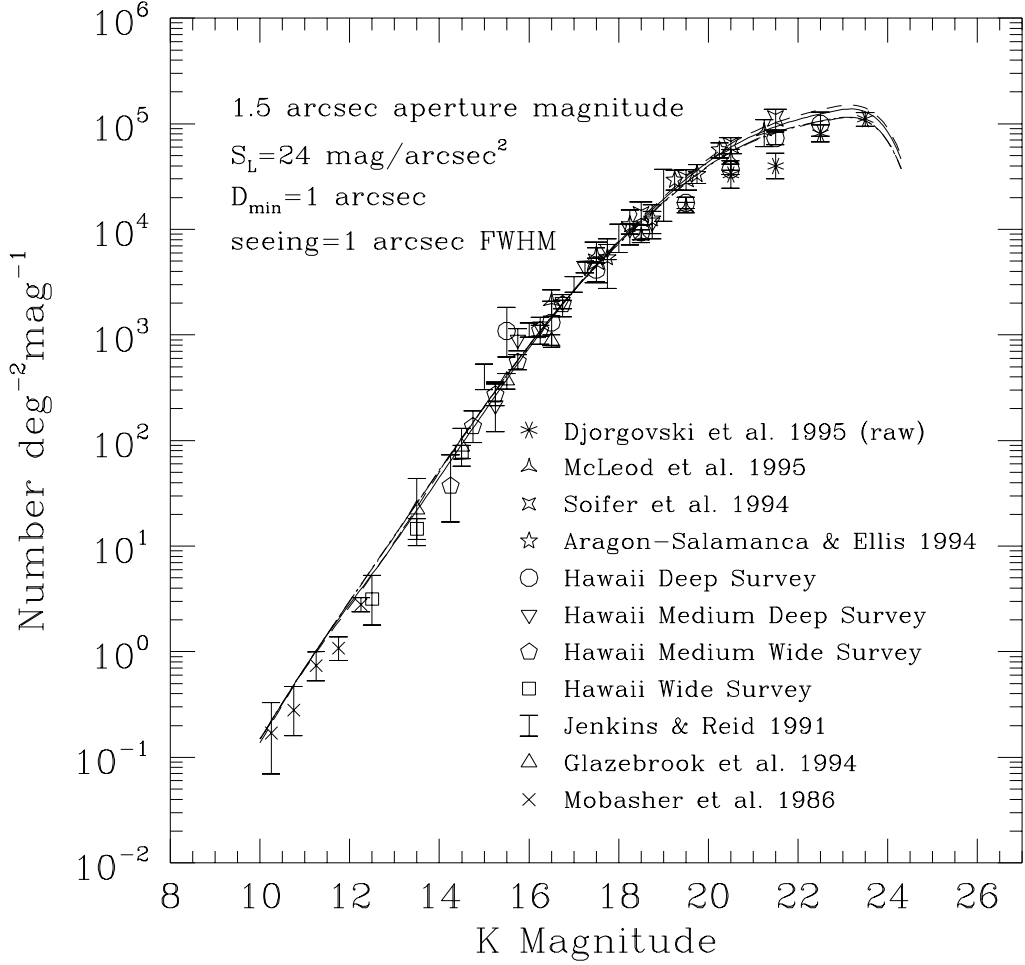

Fig. 2.— The differential number counts of galaxies in the infrared K band in a 1 mag interval in a unit area of 1 deg$^2$; the same as Figure 1, but for only a low density, flat universe ($\Omega_0 = 0.2, \lambda_0 = 0.8$, $h = 0.8$) with the four different models of evolution of spirals; S1 (thick line), S2 (short dashed line), I1 (dot and short-dashed line), and I2 (dot and long-dashed line). The standard SFR model for elliptical galaxies is used in common.

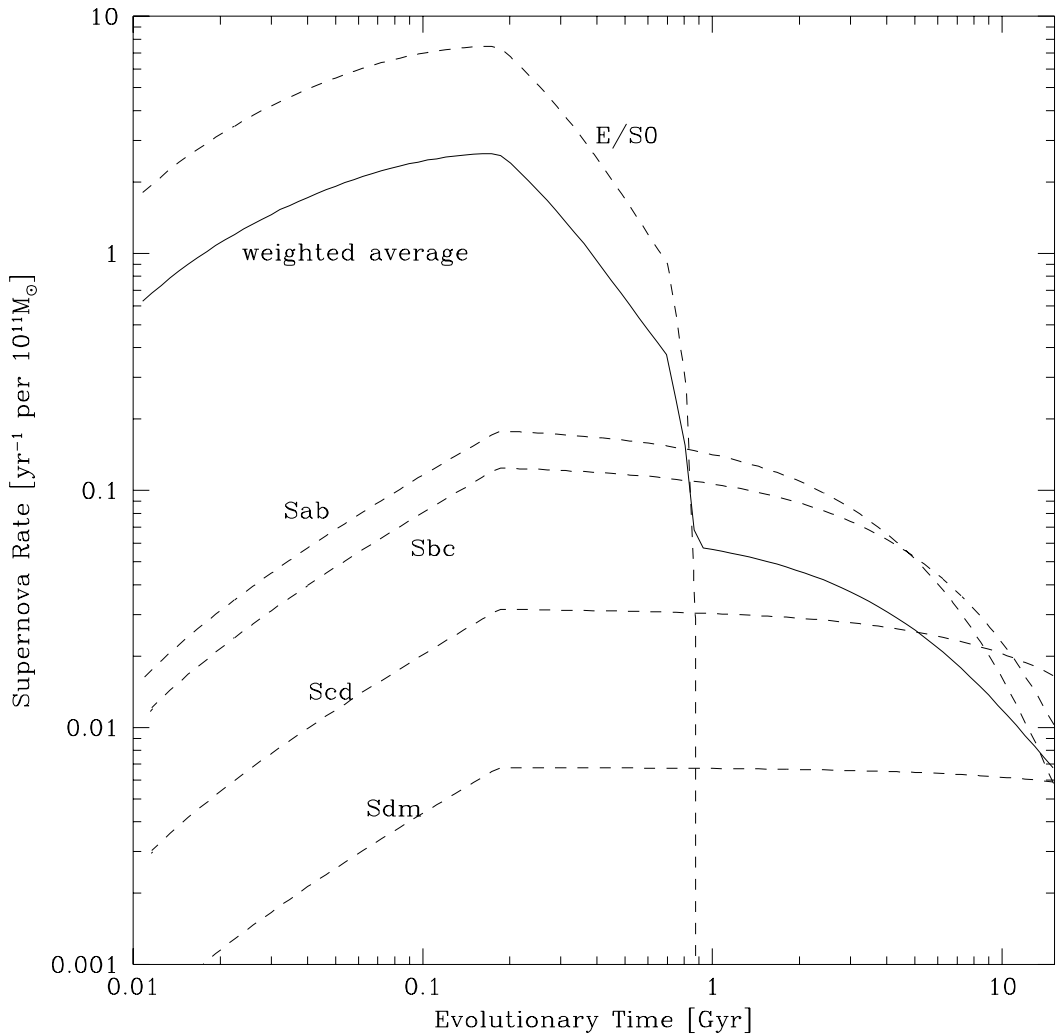

Fig. 3.— Time variation of supernova rates in a galaxy with the mass of $10^{11} M_\odot$ for various galaxy types. The time is elapsed from the epoch of galaxy formation. The dashed lines show the rates for the individual galaxy types, and the thick line for the weighted average over the types (see text). The S1 model is used to represent the evolution of spiral galaxies, and the standard SFR model for elliptical galaxies.

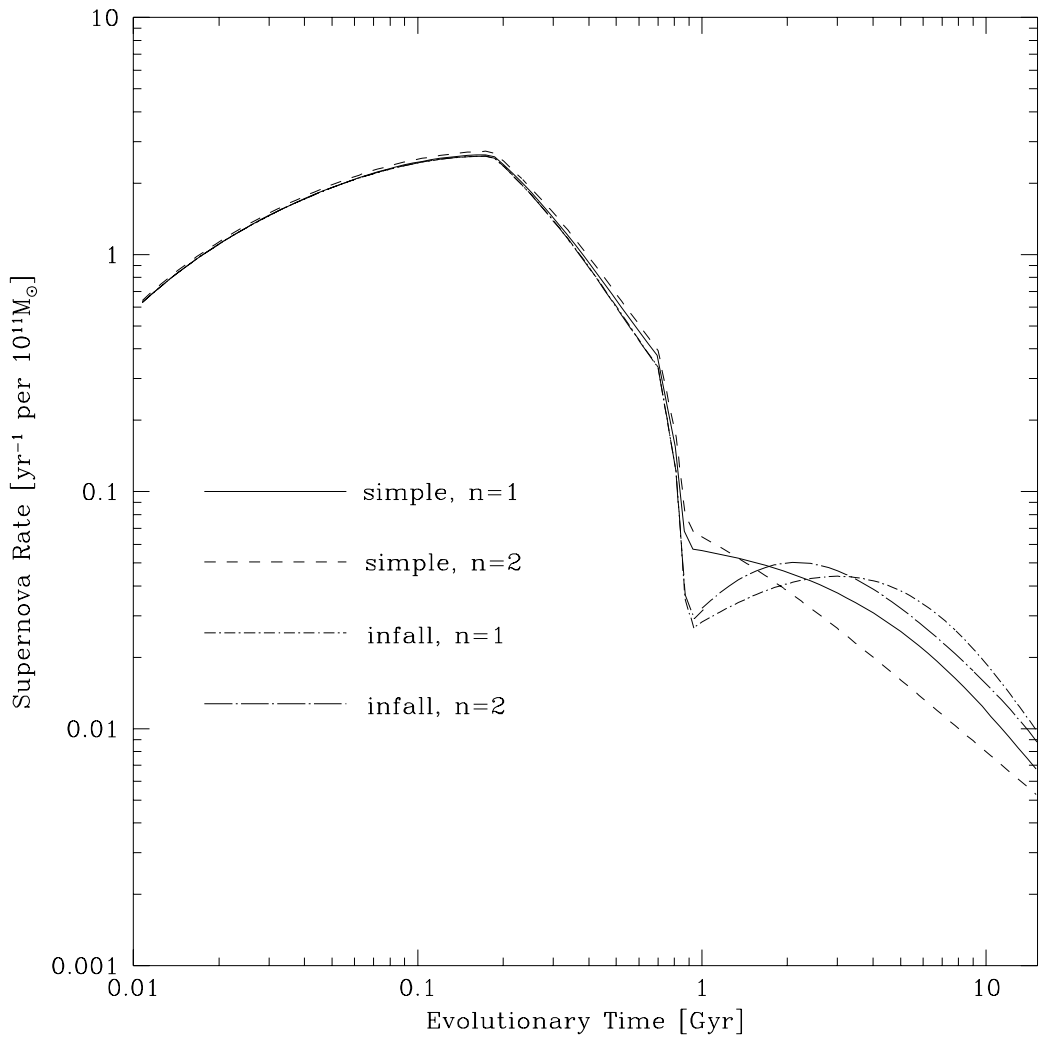

Fig. 4.— Time variation of supernova rates in a galaxy with the mass of $10^{11} M_\odot$; the same as Figure 3 but for only the weighted average rates with four different evolution models of spiral galaxies; S1 (thick line), S2 (short dashed line), I1 (dot and short-dashed line), and I2 (dot and long-dashed line). The standard SFR model for elliptical galaxies is used in common.

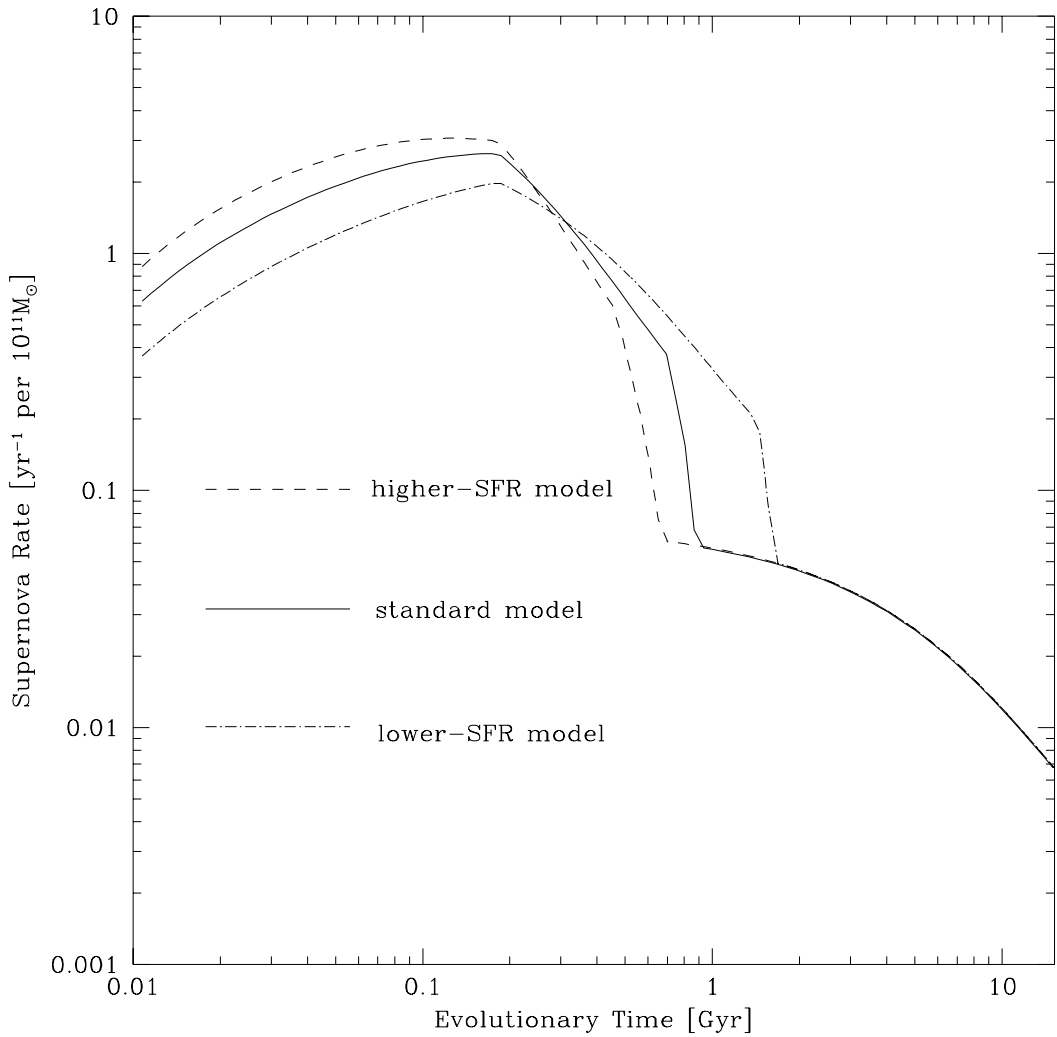

Fig. 5.— There is shown the time variation of supernova rates in a galaxy with the mass of $10^{11} M_\odot$; the same as Figure 3 but for only the weighted average rates for three different evolution models of elliptical galaxies; the lower SFR (dot and short-dashed line), standard SFR (thick line), and higher SFR (short-dashed line) models. The S1 model for spiral galaxies is used in common.

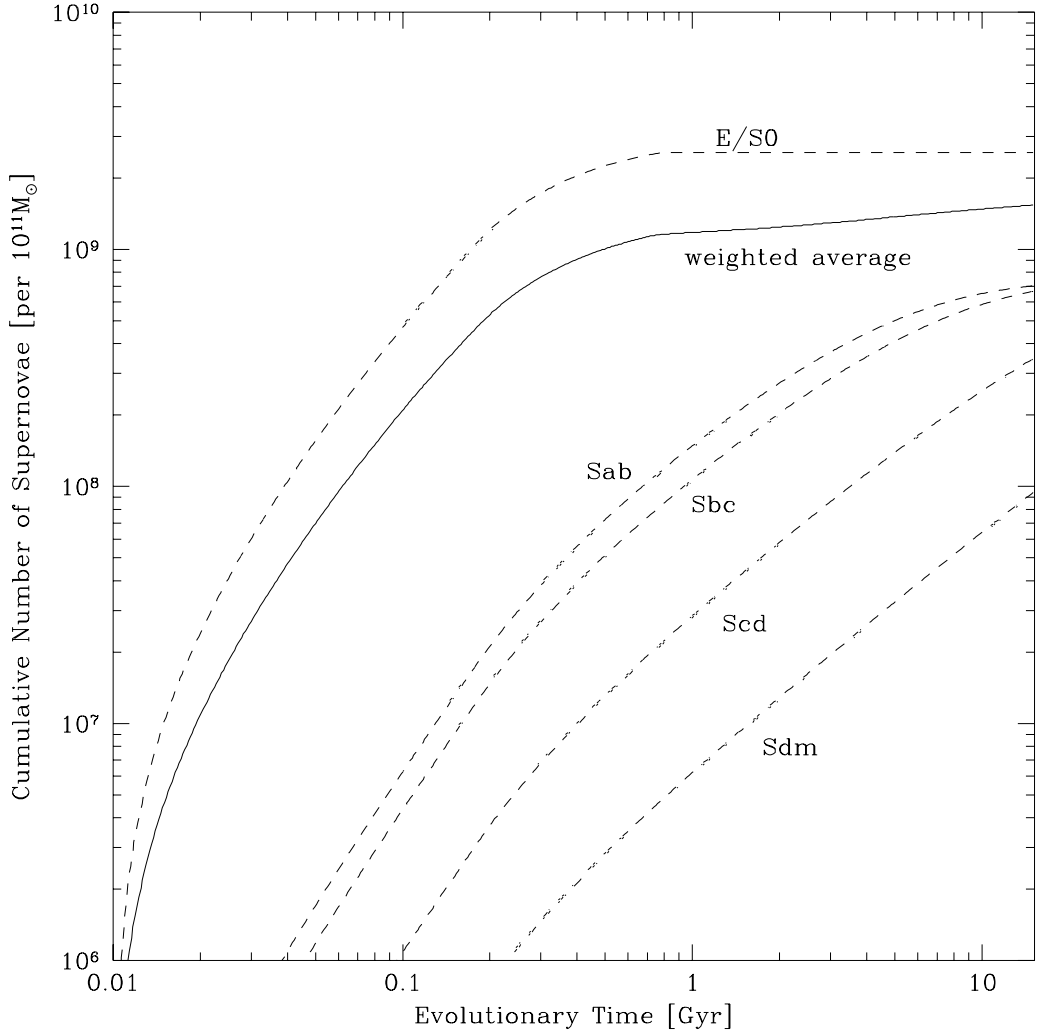

Fig. 6.— Cumulative numbers of supernovae in a galaxy with the mass of $10^{11} M_\odot$ as a function of time elapsed from the epoch of galaxy formation. The dashed lines correspond to the cases for the individual galaxy types, and the thick line to the weighted average over the types. The S1 model is used to represent the evolution of spiral galaxies, and the standard SFR model for elliptical galaxies.

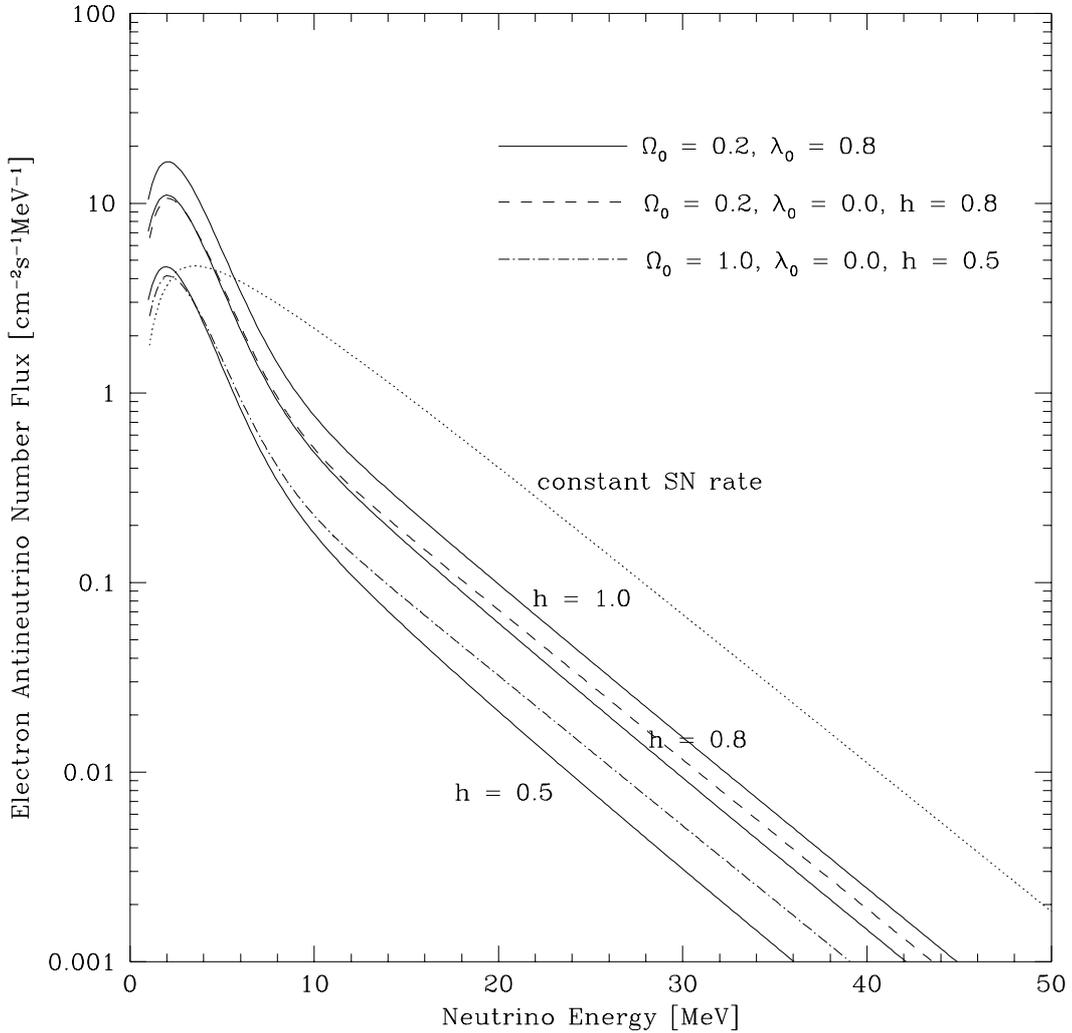

Fig. 7.— Energy spectra of the supernova relic neutrino background (SRN) for various cosmological models. All lines represent the differential number flux of electron antineutrinos ($\bar{\nu}_e$'s). The thick lines show the SRN spectrum for a low-density, flat universe with $\Omega_0 = 0.2$, $\lambda_0 = 0.8$, and three different values of the Hubble constant; $h = 0.5$ (bottom line), $h = 0.8$ (middle line), and $h = 1.0$ (top line). The dashed line shows the SRN spectrum for a low-density, open universe with $\Omega_0 = 0.2$, $\lambda_0 = 0.0$, and $h = 0.8$, and the dot-dashed line for an Einstein-de Sitter universe with $\Omega_0 = 1.0$, $\lambda_0 = 0.0$, and $h = 0.5$. The S1 model is used to represent the evolution of spiral galaxies, and the standard SFR model for elliptical galaxies. The epoch of galaxy formation is set to be $z_F = 5$. For an illustrative purpose of comparison, the SRN spectrum based on a model with a constant supernova rate is also shown by the dotted line, for which the constant supernova rate is normalized in such a way that the total (energy-integrated) SRN flux is the same with that for the model with $\Omega_0 = 0.2$, $\lambda_0 = 0.8$, and $h = 0.8$ (thick, middle line).

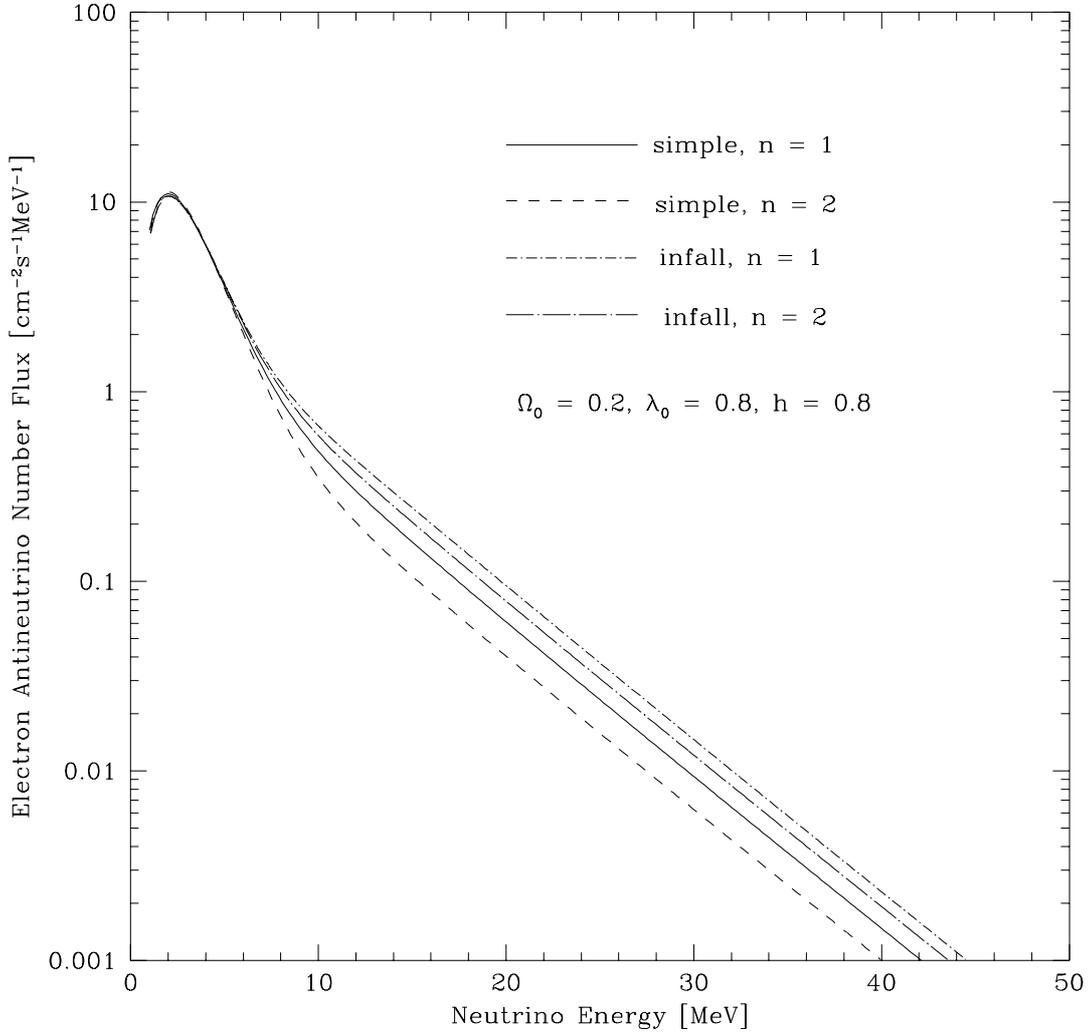

Fig. 8.— Energy spectra of the supernova relic neutrino background (SRN); the same as Figure 7 but for only a low-density, flat universe ($\Omega_0 = 0.2$, $\lambda_0 = 0.8$, $h = 0.8$) with four different evolution models of spiral galaxies; S1 (thick line), S2 (short dashed line), I1 (dot and short-dashed line), and I2 (dot and long-dashed line). The standard SFR model for elliptical galaxies is used in common, and the epoch of galaxy formation is set to be $z_F = 5$.

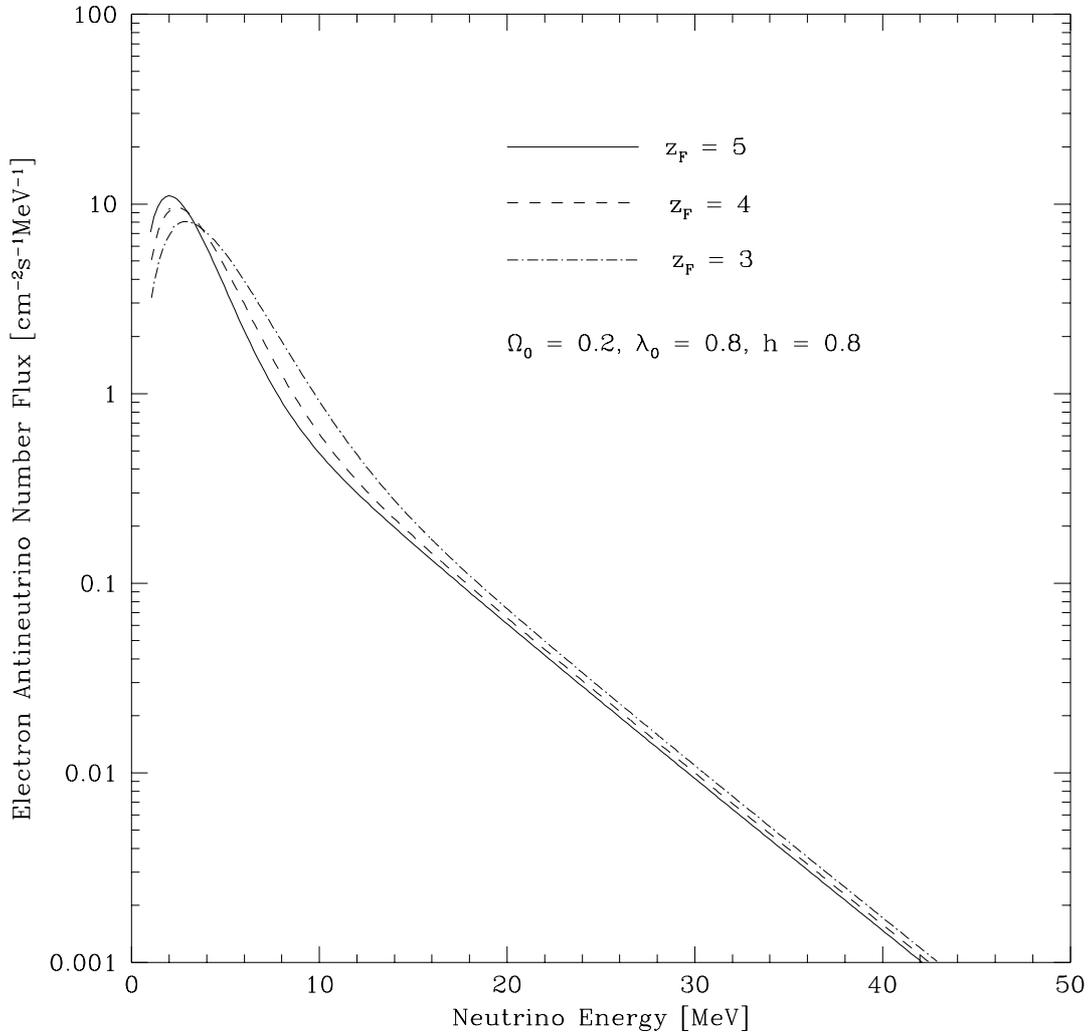

Fig. 9.— Energy spectra of the supernova relic neutrino background (SRN); the same as Figure 7 but for only a low-density, flat universe ($\Omega_0 = 0.2$, $\lambda_0 = 0.8$, $h = 0.8$) with the three different values of the epoch of galaxy formation; $z_F = 5$ (thick line), $z_F = 4$ (dashed line), and $z_F = 3$ (dot-dashed line). The S1 model is used to represent the evolution of spiral galaxies, and the standard SFR model for elliptical galaxies.

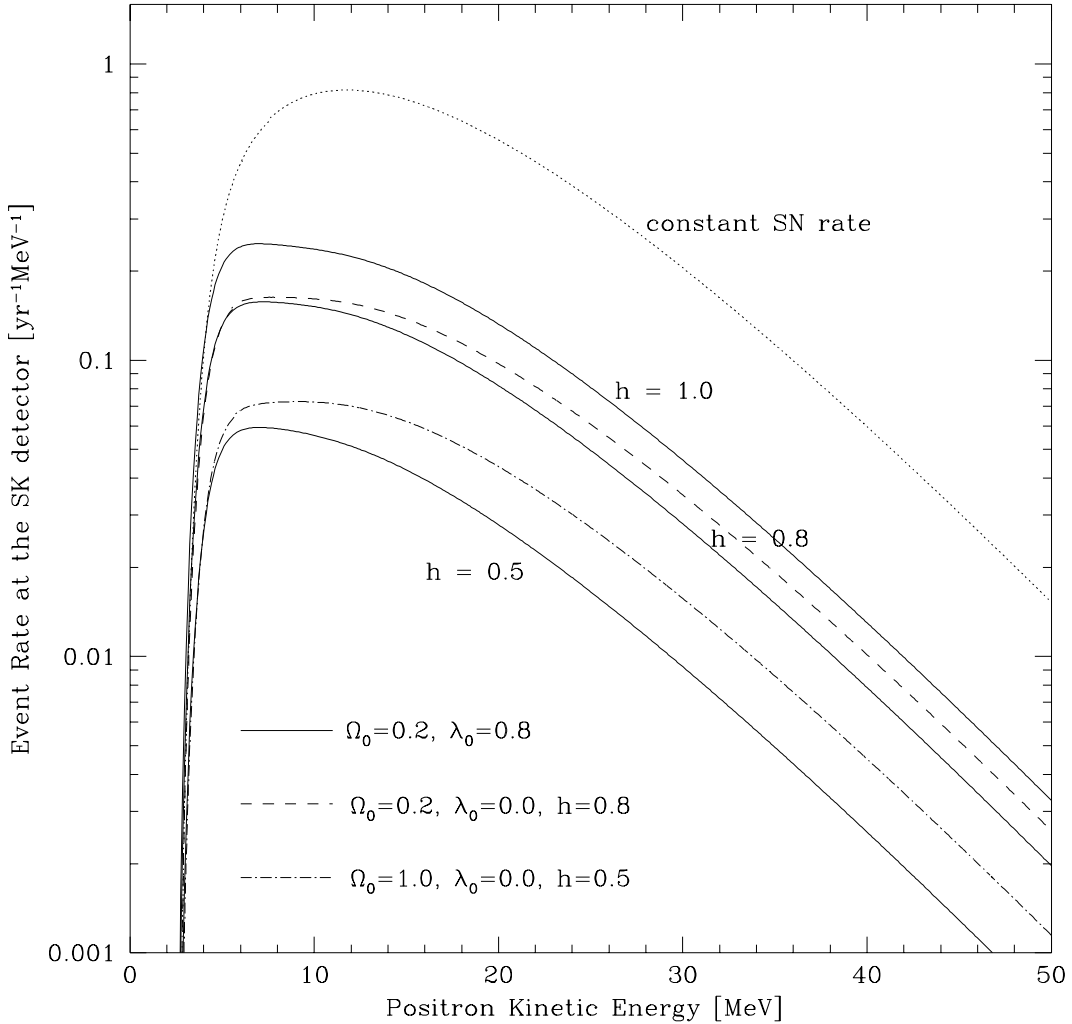

Fig. 10.— Expected event rates at the Superkamiokande (SK) detector for various cosmological models, as a function of kinetic energy of positrons produced by the reaction $\bar{\nu}_e p \to e^+ n$. This figure corresponds to Figure 7 of the SRN spectrum. The thick lines show the event rate for a low-density, flat universe with $\Omega_0 = 0.2$, $\lambda_0 = 0.8$, and three different values of the Hubble constant; $h = 0.5$ (bottom line), $h = 0.8$ (middle line), and $h = 1.0$ (top line). The dashed line shows the event rate for a low-density, open universe with $\Omega_0 = 0.2$, $\lambda_0 = 0.0$, and $h = 0.8$, and the dot-dashed line for an Einstein-de Sitter universe with $\Omega_0 = 1.0$, $\lambda_0 = 0.0$, and $h = 0.5$. The S1 model is used to represent the evolution of spiral galaxies, and the standard SFR model for elliptical galaxies. The epoch of galaxy formation is set to be $z_F = 5$. For an illustrative purpose of comparison, the event rate based on a model with a constant supernova rate is also shown by the dotted line, for which the constant supernova rate is normalized in such a way that the total SRN flux is the same with that for the model with $\Omega_0 = 0.2$, $\lambda_0 = 0.8$, and $h = 0.8$ (thick, middle line).

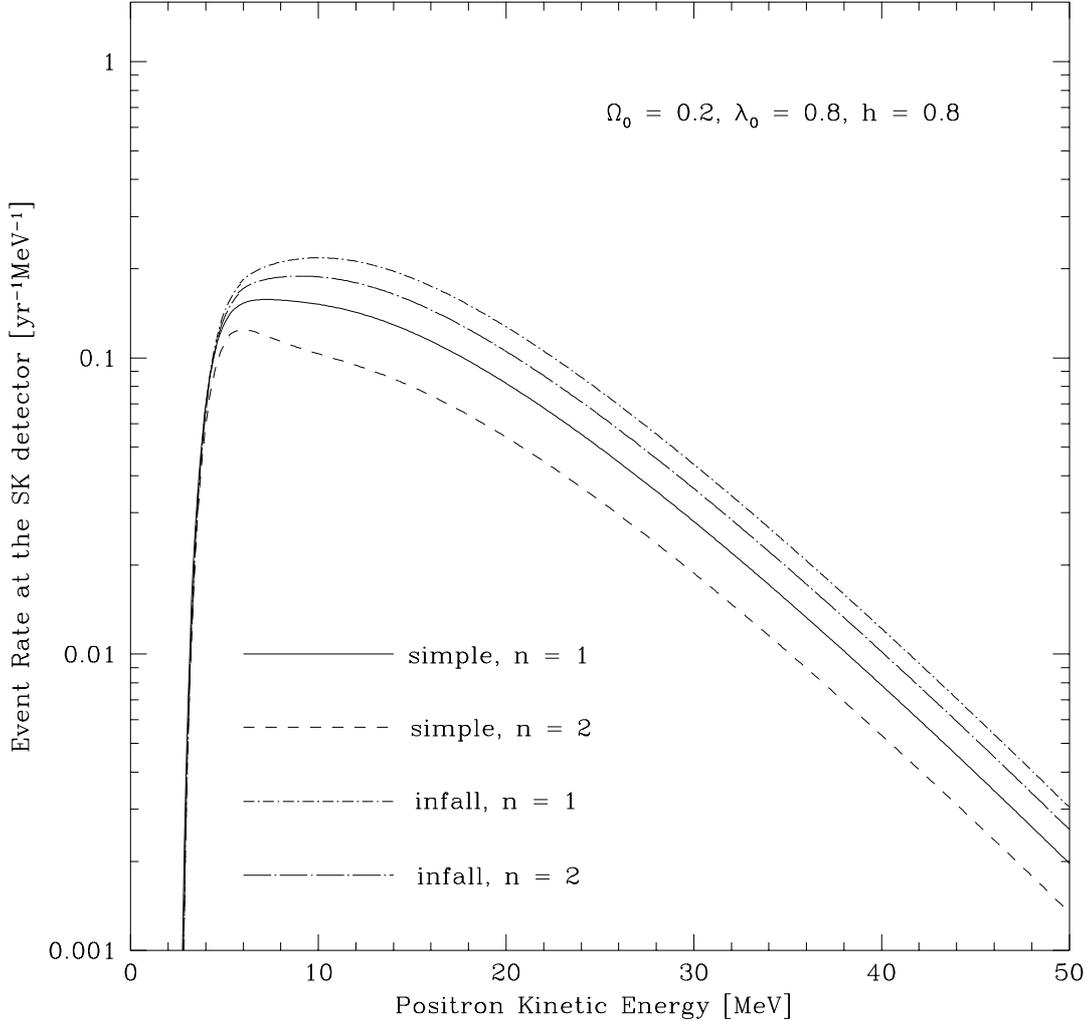

Fig. 11.— Expected event rates at the Superkamiokande (SK) detector; the same as Figure 10, but for only a low-density, flat universe ($\Omega_0 = 0.2$, $\lambda_0 = 0.8$, $h = 0.8$) with four different evolution models of spiral galaxies; S1 (thick line), S2 (short dashed line), I1 (dot and short-dashed line), and I2 (dot and long-dashed line). This figure corresponds to Figure 8 of the SRN spectrum. The standard SFR model for elliptical galaxies is used in common, and the epoch of galaxy formation is set to be $z_F = 5$.